\documentclass[twocolumn,preprintnumbers,amsmath,amssymb,prl,superscriptaddress]{revtex4-1}
\usepackage[T1]{fontenc}
\usepackage[latin1]{inputenc}
\usepackage{graphicx}

\expandafter\let\csname equation*\endcsname\relax
\expandafter\let\csname endequation*\endcsname\relax
\usepackage{amsmath}
\usepackage{xcolor}

\newcommand{\etal}{\mbox{\emph{et al.}\hspace{-1pt}}~}

\newcommand{\ev}{\mathrm{eV}}

\newcommand{\angstrom}{\textup{\AA}}
\newcommand{\nm}{\mathrm{nm}}
\newcommand{\cm}{\mathrm{cm}}

\newcommand{\Eqref}[1]{Eq.~(\ref{#1})}
\newcommand{\Figref}[1]{Fig.~\ref{#1}}

\makeatother

\newcommand{\tsiesta}{\textsc{TranSiesta}}
\newcommand{\tbtrans}{\textsc{TBtrans}}
\newcommand{\sisl}{\textsc{sisl}}

\newif\ifchanged
\changedtrue
\changedfalse

\begin{document}

\title{Full-scale Simulation of Electron Transport in Nanoporous Graphene: \protect\\Probing the Talbot Effect}

\author{Gaetano Calogero}
\email{gaca@nanotech.dtu.dk}
\affiliation{Dept. of Micro- and Nanotechnology, Technical University of Denmark, Center for Nanostructured Graphene (CNG), {\O}rsteds Plads, Bldg.~345E, DK-2800 Kongens Lyngby, Denmark}

\author{Nick R. Papior}
\affiliation{Dept. of Micro- and Nanotechnology, Technical University of Denmark, Center for	Nanostructured Graphene (CNG), {\O}rsteds Plads, Bldg.~345E, DK-2800 Kongens Lyngby,	Denmark}

\author{Bernhard Kretz}
\affiliation{Institute of Theoretical Physics, University of Regensburg, 93040 Regensburg, Germany}

\author{Aran Garcia-Lekue}
\affiliation{Donostia International Physics Center (DIPC), 20018 San Sebastian, Spain.}
\affiliation{Ikerbasque, Basque Foundation for Science, 48013 Bilbao, Spain.}

\author{Thomas Frederiksen}
\affiliation{Donostia International Physics Center (DIPC), 20018 San Sebastian, Spain.}
\affiliation{Ikerbasque, Basque Foundation for Science, 48013 Bilbao, Spain.}

\author{Mads Brandbyge}
\affiliation{Dept. of Micro- and Nanotechnology, Technical University of Denmark, Center for	Nanostructured Graphene (CNG), {\O}rsteds Plads, Bldg.~345E, DK-2800 Kongens Lyngby,	Denmark}
%\email{mads.brandbyge@nanotech.dtu.dk}

\date{\today}

\begin{abstract}
Designing platforms to control phase-coherence and interference of electron waves is a cornerstone for future quantum electronics, computing or sensing. Nanoporous graphene (NPG) consisting of linked graphene nanoribbons has recently been fabricated using molecular precursors and bottom-up assembly [Moreno \textit{et al}, \textit{Science} \textbf{360}, 199 (2018)] opening an avenue for controlling the electronic current in a two-dimensional material. By simulating electron transport in real-sized NPG samples we predict that electron waves injected from the tip of a scanning tunneling microscope (STM) behave similarly to photons in coupled waveguides, displaying a Talbot interference pattern. We link the origins of this effect to the band structure of the NPG and further demonstrate how this pattern may be mapped out by a second STM probe. We enable atomistic parameter-free calculations beyond the 100 nm scale by developing a new multi-scale method where first-principles density functional theory regions are seamlessly embedded into a large-scale tight-binding.
\end{abstract}


\maketitle


\paragraph{\textbf{Introduction}}
Controlling electron waves by harnessing phase-coherence and interference effects is a cornerstone for future nano-electronics \cite{Chen2018, Biele2017}, quantum computing \cite{Karra2016}, sensing \cite{Prindle2011} or electron beam splitting \cite{Brandimarte2017}. To this end, design of platforms with well-defined, narrow and low-loss propagation channels is essential.

Nanoporous graphene (NPG) holds great potential for distributing and controlling currents at the nanoscale \cite{Pedersen2008, Bai2010}. 
By achieving bottom-up synthesis and transfer of atomically precise NPGs Moreno \etal \cite{Moreno2018} have recently paved the way for fabrication of high-quality NPG-based devices.
Further functionalization or engineering of the pore edges could offer additional opportunities to manipulate electron transport \cite{Caridad2018}.
The particular edge topology of the linked graphene nanoribbons (GNRs) making up the
NPG results in a pronounced in-plane anisotropy, which is reflected in the
electronic structure as a peculiar energy-dependent 1D localization of electron states near the conduction band. 
Two-terminal electrical measurements and simulations successfully proved the semiconducting nature of NPG and the anisotropic electron propagation within the mesh \cite{Moreno2018}.
However, the fixed electrode setup did not answer the question of how transmitted electrons are confined in a single GNR and whether one-dimensional directed electron flow is possible. Moreover, while there is a clear analogy between the behavior of electrons in graphene and photons in dielectrics, and inspiration from optics has been drawn for graphene-based devices \cite{Chen2016, Caridad2016, Boggild2017, Xu2018}, it is not clear whether the optical analogy is applicable for NPGs with feature sizes smaller than 1 nm.

Visualizing current flow in NPG represents the most direct route to tackle these questions in an experiment. Several techniques based on superconducting interferometry \cite{Allen2015, Silveiro2015}, scanning gate microscopy \cite{Bhandari2016} or diamond-NV centers \cite{Tetienne2017} have demonstrated how imaging real-space current flow in graphene structures can profitably underpin standard electrical measurements for classical or quantum transport phenomena, while scanning probe spectroscopies are rapidly proving to be able to probe these currents at the atomic scale \cite{Kolmer2017}. 

In this article we theoretically investigate to what extent one can inject currents along individual
GNR channels in gated NPG-based devices contacted chemically to
a scanning tunneling microscope (STM) probe. We develop a multi-scale method based on Density Functional Theory
(DFT) and Non-Equilibrium Green's functions (NEGF), enabling current calculations with devices longer than 100 nm, relevant for experiments. This is accomplished by linking a perturbed contact region described by
DFT to an unperturbed large-scale region described by an effective tight-binding (TB)
model parametrized from DFT.  We find that the inter-GNR coupling disrupts the longitudinal
electron confinement into individual channels, giving rise to the Talbot effect, a
fascinating interference phenomenon predicted to occur in discrete optical wave-guide systems
\cite{Talbot1836, Berry2001, Iwanow2005, Chen2015, Walls2016, Wang2017}. 
The fine detail of Talbot wave interference is the origin of various technological applications, ranging from lithography \cite{Solak2011} to phase-contrast interferometry \cite{Birnbacher2016}. Besides, it was predicted to occur for massless Dirac Fermions in graphene \cite{Walls2016} as well as plasmons in single-mode GNRs arrays \cite{Wang2017}. 
We prove the robustness of the electronic Talbot effect in NPG by injecting current from the STM tip into various NPG sites and under various gating conditions. Using proof-of-principle calculations we also predict how this effect can be detected using a second STM tip.

\begin{figure}
	\centering
	\includegraphics[width=1.0\columnwidth]{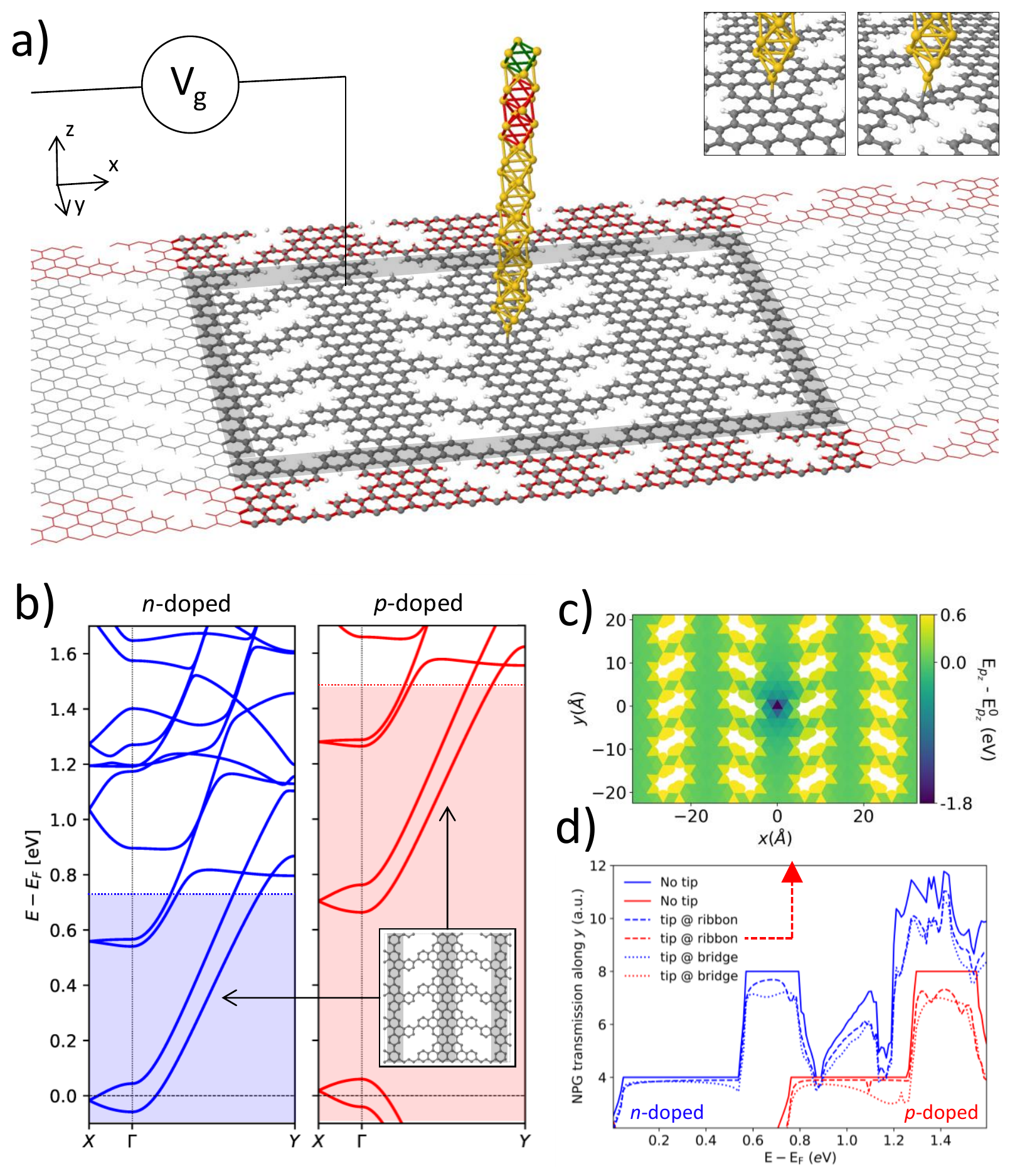}
	\caption{\label{fig:0} \textbf{DFT model of NPG.}  (a) DFT model of gated
            NPG-based device with contact to STM tip. Electrodes are highlighted
            in red. Insets show the optimized contact regions for ribbon and
            bridge positions. The shaded area delimits the device region.
            (b) band structure of the NPG unit cell in two gated conditions inducing
            either $n$-type (blue) or $p$-type (red) doping. States with energies in the shaded areas, i.e. up to $\sim0.7\,\ev$ above (below) the conduction (valence) band, disperse more along $\Gamma-Y$ than along $\Gamma-X$, hence defining predominantly longitudinal transport channels.
            (c) On-site potential of C $p_z$ orbitals in a $p$-doped NPG. Values are relative to the average potential $E^0_{p_z}$ in the bulk of ribbons $\sim 0.5\,\nm$ away from the tip. Beyond this distance the perturbation from the tip is effectively screened.
            (d) Transmission between the two NPG electrodes, with and without tip contact,
            for the two different tip positions in (a) and gated conditions in (b). }
\end{figure}

\paragraph{\textbf{Multi-scale method}}
We carry out transport calculations based on NEGF \cite{Brandbyge2002, Datta2000} applied to DFT or TB models, using \tsiesta, \tbtrans\ \cite{Soler2002, Papior2017} and \sisl\ \cite{Papior2017a}. While parameter-free DFT models limit the accessible device sizes to only a few nanometers, parameterized TB can capture basic transport features of large systems with minor chemical perturbations (e.g. metal contact or chemical
hybridization).
The DFT model for the NPG device, shown in \Figref{fig:0}a, covers an area of $6.5\times5.5\,\nm^2$ (1449 atoms). It is defined on a single-$\zeta$ polarized basis set
\footnote{This choice guarantees enough accuracy for the purposes of this work. A complete description of the NPG electronic structure would requires more complex basis sets \cite{Moreno2018, Papior2018}.} 
and with open (periodic) boundary conditions along
$y$ ($x$). Pores are passivated with H atoms and a model Au tip is in chemical contact with the NPG, with Au-C bond-length of $\sim2.0\,\angstrom$. In order to describe a more realistic experimental environment we include a bottom gate in the calculations. This is done by fixing spatial charge in a \emph{gate-layer} placed $15\,\angstrom$ beneath the NPG, so as to dope it by $\pm 10^{13}\,e^{-}/\cm^2$ \cite{Papior2015}.
The device Green's function is calculated as
\begin{equation}\label{eq:negf}
\mathbf G = \left[ \mathbf S\,(E + i\eta) - \mathbf H - \sum_{i}\boldsymbol\Sigma_i \right]^{-1}
\end{equation}
where $\mathbf S$ and $\mathbf H$ are the model overlap and Hamiltonian and $\boldsymbol\Sigma_i$, called \emph{self-energies}, connect the device to
semi-infinite regions along $x$ (electrodes).

A self-energy allows to seamlessly connect different regions, be it infinite bulk or localized perturbations. We exploit this to embed one or more DFT-precision regions inside much larger TB models, needed to reach experimental relevant dimensions and observe interference. The central idea of this multi-scale approach is to construct the self-energy connecting $p_z$ orbitals of the outermost unperturbed DFT atoms (shaded frame in \Figref{fig:0}a) to a larger $p_z$ TB model. We obtain TB parameters from a DFT calculation of unperturbed NPG, such that the resulting model retains the interaction range of the DFT basis set, is non-orthogonal and takes into account self-consistent effects of gates and/or bias. Therefore, besides introducing local DFT-precision in the TB model, this method enables TB-based $N$-electrode transport calculations without any fitting parameters (see Supporting Information for further details).
\paragraph{\textbf{DFT calculations}}

The positions of the outermost tip atom and the nearby $\sim20$ C atoms in the three-probe device are optimized until forces are less than $0.01\,\ev/\angstrom$. We consider two different positions of the tip, namely on top of a C atom in the middle of a ribbon or at a bridge between two C atoms linking two ribbons (see inset to \Figref{fig:0}a). In the former case the C atoms below the tip are pushed $\sim0.3\,\angstrom$ below the NPG
plane, while at the bridge site the tip binds to two aromatic rings causing a slight torsion.

The main effect of gating on the NPG electronic structure is a rigid band shift of $\pm 0.35\,\ev$ for $n$-
and $p$-type doping cases, respectively (\Figref{fig:0}b). We find in both cases a band gap of $\sim 0.7\,\ev$ between symmetric valence and conduction bands, in good agreement with results obtained for non-gated NPG \cite{Moreno2018}. Bands with predominant longitudinal character are clearly visible at energies up to $0.7-0.8\,\ev$ above (below) the conduction (valence) band (shaded areas), where states along $\Gamma-Y$ indeed disperse more than those along $\Gamma-X$.
The contact with the tip causes local variation of the C $p_z$ potential which is screened at $\sim0.5\,\nm$ from the tip (\Figref{fig:0}c). 
This degrades transmission between the two NPG electrodes along the $y$ direction (\Figref{fig:0}d). For all energies in the longitudinal regime transport is essentially one-dimensional along the GNRs, due to the weak inter-ribbon coupling. 
The onset at $E\sim 0.9\,\ev$ of bands
with dominant transverse character disrupts the 1D confinement and conductance
quantization is lost. 
For all the considered gating conditions and tip positions we find qualitatively similar electronic and transport features (see Supporting Information). Below without loss of generality we mainly focus on $n$-doped NPG probed at a ribbon site, highlighting differences only where relevant.

\paragraph{\textbf{Multi-scale calculations}}
\begin{figure}
	\centering
	\includegraphics[width=1.0\columnwidth]{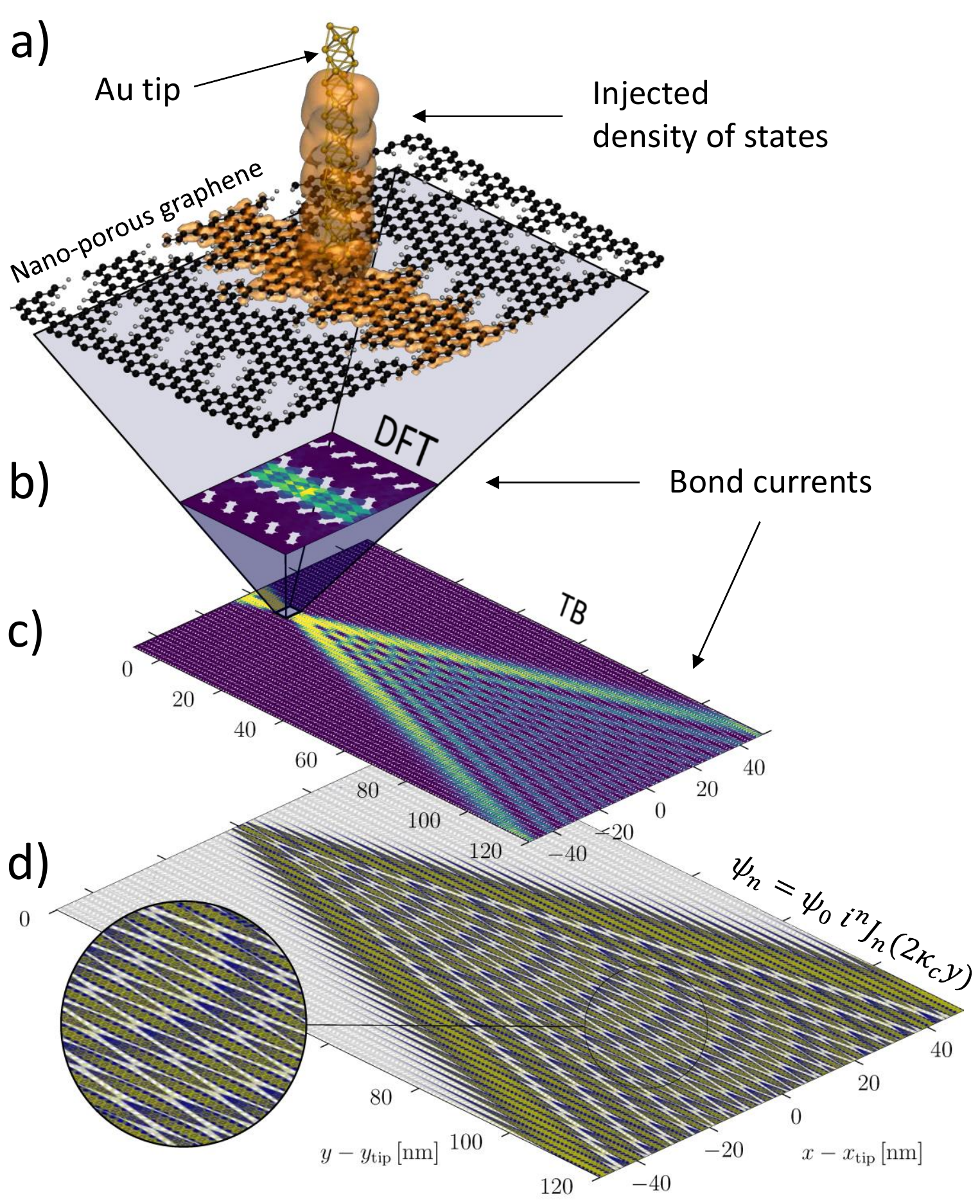}
	\caption{\label{fig:1}
		\textbf{Talbot effect from multi-scale calculations.}
		The figure shows (a) DFT geometry and density of states injected by the tip, (b) bond-currents at $E-E_{\mathrm F}=0.2\,\ev$ obtained with the DFT model and (c) a large-scale TB model with DFT-precision injection, and (d) the best-fit to the Talbot effect equations.
		In (d) the local spectral density of states (ADOS) from the tip is shown in yellow, scaled by $y-y_{\mathrm{\,DFT\, tip}}$ to compensate damping occurring far from the tip. The fitted equations $\psi_n$ are plotted in blue underneath the ADOS.
		The integers $n\in[-26,26]$ index the $53$ GNRs assembling the $100\,\nm \times 120\,\nm$ NPG.
	}
\end{figure}
We visualize the electron flow near the contact through the injected density of states\footnote{Injected density of states is calculated by summing the absolute modules of the three eigenchannels \cite{Paulsson2007} which contribute the most to transmission from the tip to the NPG electrodes.} 
and bond-currents from the tip,\footnote{Bond-currents are defined as \unexpanded{$J_{\alpha\beta} = \sum_{\nu\in\alpha}\sum_{\mu\in\beta} J_{\nu\mu}$}, where \unexpanded{$\nu$} (\unexpanded{$\mu$}) indicates basis orbitals centered on atom \unexpanded{$\alpha$} (\unexpanded{$\beta$}).} 
as shown in \Figref{fig:1}a-b. These clearly demonstrate electron confinement inside the probed ribbon up to distances comparable to the DFT cell size, for all energies ranging from the conduction band up to $\sim0.7\,\ev$ above it (see Supporting Information).
We study the distribution of electronic current in the far-field, i.e. far from the source by performing transport calculations where the DFT
``injection region'' is embedded into a larger TB NPG model. 
We benchmark the
applicability of this method by embedding the DFT-precision injection region in a
TB-region which has the same boundary conditions and size of the DFT transport setup (see Supporting Information). The results show that we can reproduce the DFT transmission spectra in the longitudinal regime (\Figref{fig:0}b, shaded) by the combined DFT+TB model.
Next we embed the DFT injection region as electrode in a large $100\,\nm\times120\,\nm$ TB model of a device with two NPG electrodes along $y$ and a complex absorbing potential \cite{Calogero2018, Xie2014} along $x$.
This larger model reveals that at distances beyond the DFT cell size transversal losses significantly affect the far-field behavior (\Figref{fig:1}c). Current splits into neighboring ribbons with a certain periodicity of $\sim7-8\,\nm$. The resulting ``beams'' diverge from the $y$ direction with a maximum angle which varies slightly with energy (see Supporting Information). In particular when the tip injects into a ribbon site this angle decreases from $\sim 30^\circ$ for $E-E_{\mathrm F}< 0.3\,\ev$ to $\sim 20^\circ$ for $0.4\,\ev < E-E_{\mathrm F} < 0.8\,\ev$. We observe similar results for injection into a bridge site, although the interference in this case is more blurred, since the injected currents start out by propagation in the two bridged ribbons (see Supporting Information).

\begin{figure}
	\centering
	\includegraphics[width=1.0\columnwidth]{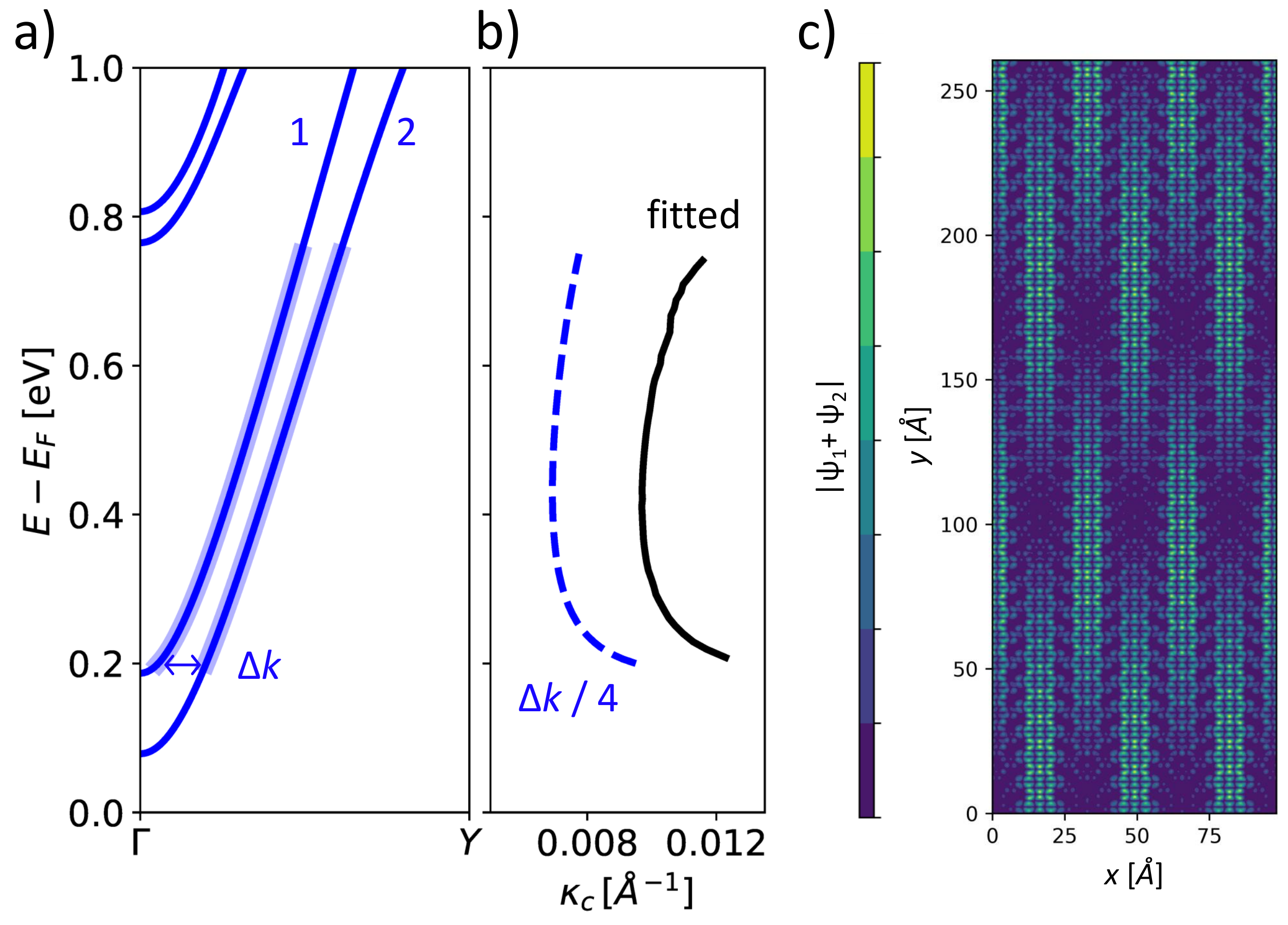}
	\caption{\label{fig:3}
		\textbf{Origin of Talbot effect}
		(a) Longitudinal bands for a $p_z$ TB model of $n$-doped NPG parameterized from DFT. 
		(b) Energy dependence of inter-GNR coupling $\kappa_c$ from fit to the Talbot equations \Eqref{eq:2}, in comparison to $\Delta k /4 = |k_{2} - k_{1}|/4$.
		(c) Amplitude of $\psi_{\mathrm{1}}+\psi_{\mathrm{2}}$ at $E-E_{\rm F}=0.2\,\ev$, showing the interference underlying the Talbot effect.
	}
\end{figure}

\paragraph{\textbf{The Talbot effect}}
The Talbot interference generally refers to repeated self-imaging of a diffraction grating \cite{Berry2001}. In this context the wavefunction $\psi_n$ inside the $n^{\mathrm{th}}$ element of an infinite array of weakly coupled discrete channels aligned along $y$ obeys the following discrete
differential equation \cite{Yariv1984, Iwanow2005, Pertsch2002, Somekh1973}
\begin{equation}\label{eq:diff}
i\,\frac{d\psi_n}{dy}(y) + \kappa_c\,\left[ \psi_{n-1}(y) + \psi_{n+1}(y) \right] = 0, 
\end{equation}
where $\kappa_c$ represents the inter-channel coupling coefficient. The coefficient
$\kappa_c$ can be regarded as a figure-of-merit for the degree of 1D confinement in the
elements of the array: the lower the value of $\kappa_c$, the lower the spread into
the weakly coupled adjacent channels.  
In the particular case when only a single channel is initially excited,
i.e. $\psi_{n=0} = \psi_0$, the solutions for \Eqref{eq:diff} can be written as
\cite{Yariv1984, Wang2017, Somekh1973}
\begin{equation}\label{eq:2}
\psi_n (y) = \psi_0 \, i^n \, J_n(2\kappa_c y), 
\end{equation}
where $J_n$ is the Bessel function of the $n^{\mathrm{th}}$ order.  

We find that the square modulus of \Eqref{eq:2} can be fitted to the far-field spectral
density of states originating from the tip.  The best fit on a set of
$53$-channels of a $100\,\nm \times120\,\nm$ $n$-doped NPG is illustrated in
\Figref{fig:1}d. The fitted values at $E=0.2\,\ev$ are $\psi_0=0.037$ and
$\kappa_c=0.012\,\angstrom^{-1}$. The latter varies slightly with energy as shown in \Figref{fig:3}b. 
The Talbot effect originates due to interference of the two longitudinal Bloch states $\psi_1$ and $\psi_2$. From the momentum difference one can estimate the coupling strength as $\kappa_c = \Delta k /4 = |k_{2} - k_{1}|/4$ \cite{Walls2016, Wang2017}. In \Figref{fig:3}c we plot $|\psi_{\mathrm{1}}+\psi_{\mathrm{2}}|$
for the NPG without tip contact, showing the regular interference 
pattern expected when all GNRs are simultaneously excited \cite{Walls2016}.

\paragraph{\textbf{Dual-probe multi-scale calculations}}

\begin{figure}
	\centering
	\includegraphics[width=1.0\columnwidth]{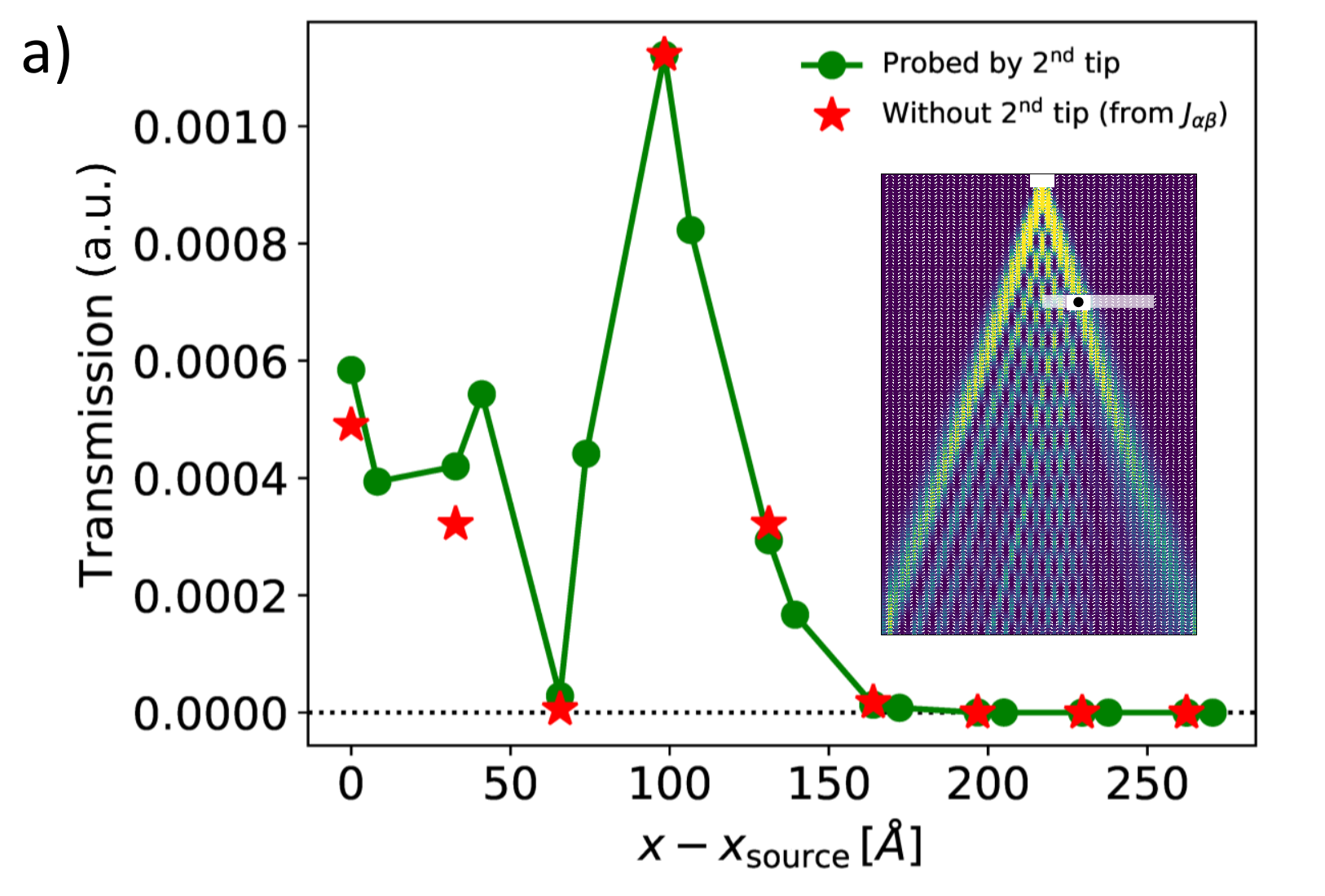}
	\caption{\label{fig:2}
		\textbf{Probing the Talbot effect.}
		(a) Transmission at $E-E_{\mathrm F}=0.2\,\ev$ (green) measured by a $2^{\mathrm{nd}}$ DFT-precision tip probing ribbon and bridge sites of $n$-doped NPG
                along the white line shown in inset, in comparison with
                bond-currents flowing in absence of the $2^{\mathrm{nd}}$ tip (red). These are obtained on a ``per ribbon'' basis by summing all bond-currents passing through the
                white line, without distinction between ribbon and bridge sites, and then
                scaling by a factor $1/16$. 
                The inset shows injected currents scattering off the $2^{\mathrm{nd}}$ tip for one of the scanned positions, indicated by the black dot.
            }
\end{figure}

The latest developments in STM have enabled measurements with up to four tips \cite{Baringhaus2016} and control over tip-tip distance down to tens of nm \cite{Kolmer2017}. 
Within this context, we propose to use dual-probe STM to prove the electronic Talbot effect experimentally.
One probe can be used to inject current into the NPG at a fixed position as discussed above while the second probe is used to map out the interference pattern.
We use the modular capability of the multi-scale approach to embed two DFT-precision tips in
the large TB model. Both tips are located
$\sim 2\,\angstrom$ above NPG and we fix the probe-tip distance to around $30\,\nm$
from the source. The probe-tip is moved between the bridge and ribbon positions along the white line shown in the inset in \Figref{fig:2}a. The calculated transmission between the two tips, injecting electrons at $E-E_{\mathrm F}=0.2\,\ev$ into $n$-doped NPG, is shown in \Figref{fig:2}a. The maxima of transmission reproduce the maxima of bond-currents injected from the source in absence of the second tip within 5\%. This proves that the signal in electric current detected by the second tip can map out the Talbot interference
pattern.

\paragraph{\textbf{Conclusions and outlook}}
In conclusion, we have explored how electrons injected by a STM probe in chemical contact to gated NPG-based devices behave in near and far field. By performing multi-scale parameter-free calculations of large-scale TB models of NPG with DFT-precision regions we found a clear signature of phase-coherence of electron waves, manifested in the far field as a Talbot interference effect. 
The origin of this phenomenon is the cross-talk between longitudinal 1D channels (GNRs) making up the NPG. Using proof-of-principle multi-probe calculations we have shown this interference effect may be observed in a dual-probe STM experiment.
Further investigations could potentially shed light on the important impact of defects,
substrate or pore functionalization on the current flow.
Importantly, chemical design of the inter-ribbon bridges allows fine-tuning of the coupling strength $\kappa_c$ to improve 1D transport confinement.
The Talbot effect in elastic, phase-coherent transport may be used
to gain insights into the phase-breaking length in these structures due to various
scattering mechanisms such as electron-phonon coupling.
Finally, since topologically non-trivial states were found at the edges of chiral
GNRs \cite{Groning2018, Rizzo2018}, further studies might potentially reveal whether topological signatures could emerge in NPGs, perhaps impacting electron dynamics.

\begin{acknowledgements}
\paragraph{\textbf{Acknowledgements}}
Financial support by Villum Fonden (00013340), Danish research council (4184-00030), Spanish Ministerio de Econom{\'i}a y Competitividad (FIS2017-83780-P and MAT2016-78293-C6-4-R), UPV/EHU (IT-756-13) is gratefully acknowledged. The Center for Nanostructured Graphene (CNG) is sponsored by the Danish Research Foundation (DNRF103). We are thankful to Jos{\'e} Caridad, Pedro Brandimarte, Isaac Alc{\'o}n Rovira and Stephen Power for useful discussions. 
\end{acknowledgements}

\newpage

\section{Supporting information}
Below we provide Supporting Information on:
i) Computational details; ii) DFT potential and charge distributions in gated NPG+tip systems; iii) transmission and near-field bond-currents from tip to DFT-modeled NPG; iv) 
multiscale method applied to NPG+tip systems; v) transmission and far-field bond-currents from DFT-precision tip to TB-modeled NPG; vi) bond-current maps from dual-probe calculations.

\subsection{Computational details}
The DFT electronic structure and optimized geometry for the NPG unit cell, with alternate $A-B$ and $B-A$ bridge environment as in Moreno \etal \cite{Moreno2018}, are obtained using the \tsiesta\ code \cite{Soler2002, Papior2017}. The unit cell is orthogonal and contains 100 atoms (80 C + 20 H). We use a SZP basis set with $0.01$ Ry energy shift. This choice neglects the existence of high-energy SAMO states, usually captured by more accurate basis sets \cite{Moreno2018}. However, since the Talbot effect occurs at much lower energies, the computationally more efficient SZP basis guarantees sufficient accuracy. 
We use norm-conserving Troullier-Martins pseudopotentials with a mesh cutoff of $400$ Ry and the PBE \cite{Perdew1996} flavour of the generalized gradient approximation (GGA) for the exchange-correlation functional. 
The Brillouin zone is sampled using a $15\times 51\times 1$ Monkhorst-Pack k-point mesh. 
Structural optimization is performed using a force threshold of $0.01\,\ev/\angstrom$. 
The geometry of the tip has been optimized so as to ensure a flat density of states over the range of energies relevant for this study.

We run zero-bias DFT-NEGF calculations with \tsiesta\ for a three-probe device where the model tip, semi-infinite along the out-of-plane direction $z$, is placed $\sim 2 \angstrom$ above the NPG, in atomic contact with it above a ``ribbon'' or ``bridge'' site.
We use a $2\times7$ NPG supercell (1449 atoms, $6.5\times5.5\,\nm^2$), periodic along the transverse direction $x$ and with two semi-infinite electrodes (\emph{top} and \emph{bottom}) along the longitudinal direction $y$. Contrary to the device setup reported in \cite{Moreno2018}, here we use semiconducting NPG electrodes rather than metallic pristine graphene ones. 
We use $5\times 5\times 1$ k-points to run the NPG+tip self-consistent calculation with \tsiesta\ and use $71\times 1\times 1$ k-points to compute transmission among the three electrodes with \tbtrans. 
We set up a gate by placing a fixed charged layer $15\,\angstrom$ below NPG such as to induce an electron density $n_g=\pm 10^{13}\,\cm^{-2}$ in it. This is important to pin the Fermi level in NPG electrodes and device DFT calculations. In this way we avoid artificial potential offset at the interfaces between scattering region and electrodes, which would otherwise occur due to the tip-induced change in the NPG work-function. 
Coordinates of the tip apex and the nearest $\sim 20$ C atoms are optimized, without gate, until forces are less than $0.01\,\ev/\angstrom$.

For the large TB+NEGF we generate NPG supercells containing up to 356.160 atoms and covering areas up to $100\times120\,\nm^2$.
No periodic boundary conditions are used (i.e. \tbtrans\  $k$-points grid includes only $\Gamma$). Injectced currents are drained into semi-infinite electrodes along $y$ and absorbed along $x$ using complex absorbing potentials (CAP) \cite{Calogero2018, Xie2014}. 

We visualize current flow in the devices plotting 2D bond-current maps, summing all bond-currents values flowing out of each atom (only positive-valued bond
currents are considered). The color map is scaled in proportion to the current magnitude, so that areas with low to zero current appear in blue. Color range is normalized to the maximum value of current in the device and is often adjusted to enhance contrast.

\subsection{DFT potential and charge distributions in gated NPG+tip systems}

\begin{figure*}
	\centering
	\includegraphics[width=1.0\textwidth]{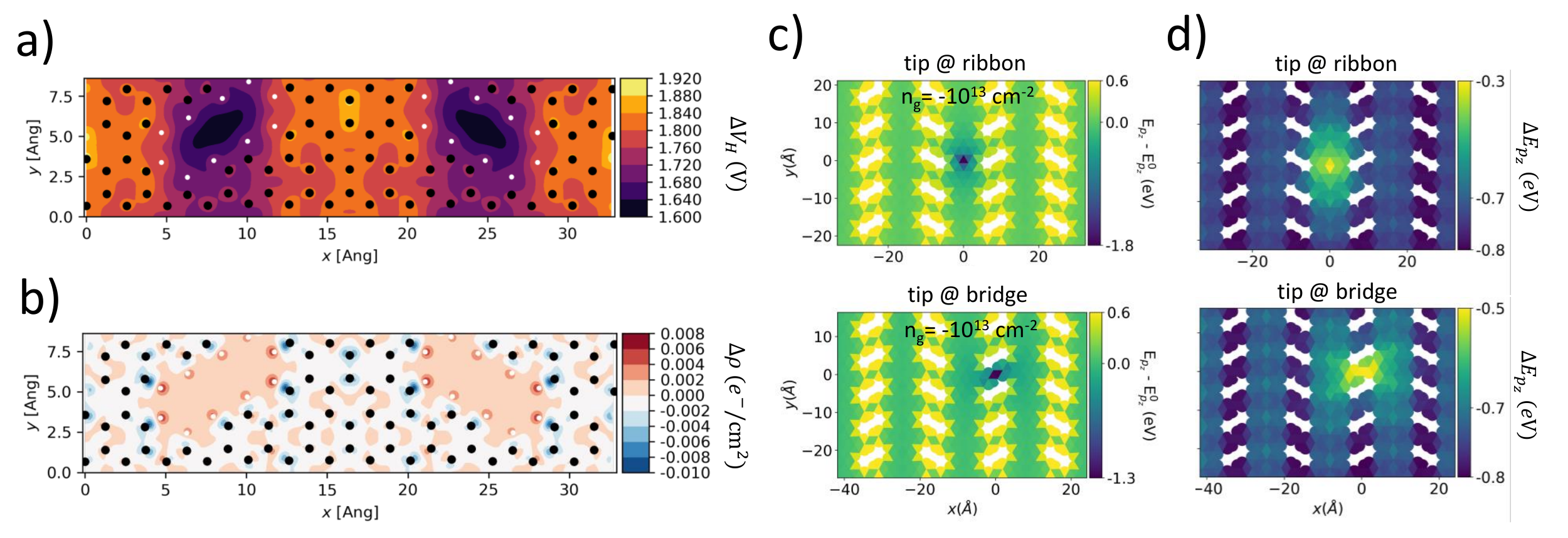}
	\caption{\label{fig:1}
		(a) Difference in electrostatic Hartree potential $\Delta V_H=V_H(n_g>0)-V_H(n_g<0)$ between the systems with positive and negative values of gate-induced electron density $n_g$.
		(b) The resulting change in charge density distribution $\Delta\rho=\rho(n_g>0)-\rho(n_g<0)$.
		(c) On-site $p_z$ potential for all C and H atoms in the three-probe device, relative to the average potential $E^0_{p_z}$ of the most internal atoms in the GNRs, showing how the gold tip dopes the NPG for $n_g=-10^{13}\,\cm^{-2}$. The regions where the maximum $p$-type and $n$-type potential shift occur are indicated in yellow and blue, respectively, for ribbon and bridge positions of the tip. 
		(d) Difference $\Delta E_{p_z}=E_{p_z}(n_g>0)-E_{p_z}(n_g<0)$ between the on-site $p_z$ potential for the n-doped and p-doped devices, for the two different tip positions.
	}
\end{figure*}

The electrostatic potential and charge distribution in the NPG are affected by application of a gate voltage as illustrated in \Figref{fig:1}a-b. Potential varies more rapidly in proximity of the narrower sections of the 7-13 GNRs. The doping charge density distribution reflects this variation, with a tendency to polarize in proximity of the regions where the potential varies the most.

Carbon $p_z$ levels are doped locally in the mesh when the gold STM tip is in contact with the NPG, as shown in \Figref{fig:1}c for $n_g=-10^{13}\,\cm^{-2}$. A maximum local $n$-doping of $\approx -1.7\,\ev$ and $\approx -1.3\,\ev$ is reached depending on contact position. For $n_g=+10^{13}\,\cm^{-2}$ the maps are very similar but the $n$-type doping is reduced by $\approx0.34\,\ev$ and $\approx0.24\,\ev$ for the two tip
positions. This is due to a higher electron population in the NPG which makes it less
favorable for electrons to transfer from the tip to the NPG. As further illustrated in \Figref{fig:1}d, regardless of the tip position or the gating we always find that the doped region in proximity of the contact region extends with a radius $\sim0.5\,\nm$.
This proves that the supercell we use is large enough to ensure an unperturbed potential in its outermost regions.

\subsection{Transmission and near-field bond-currents from tip to DFT-modeled NPG}

\begin{figure*}
	\centering
	\includegraphics[width=1.0\textwidth]{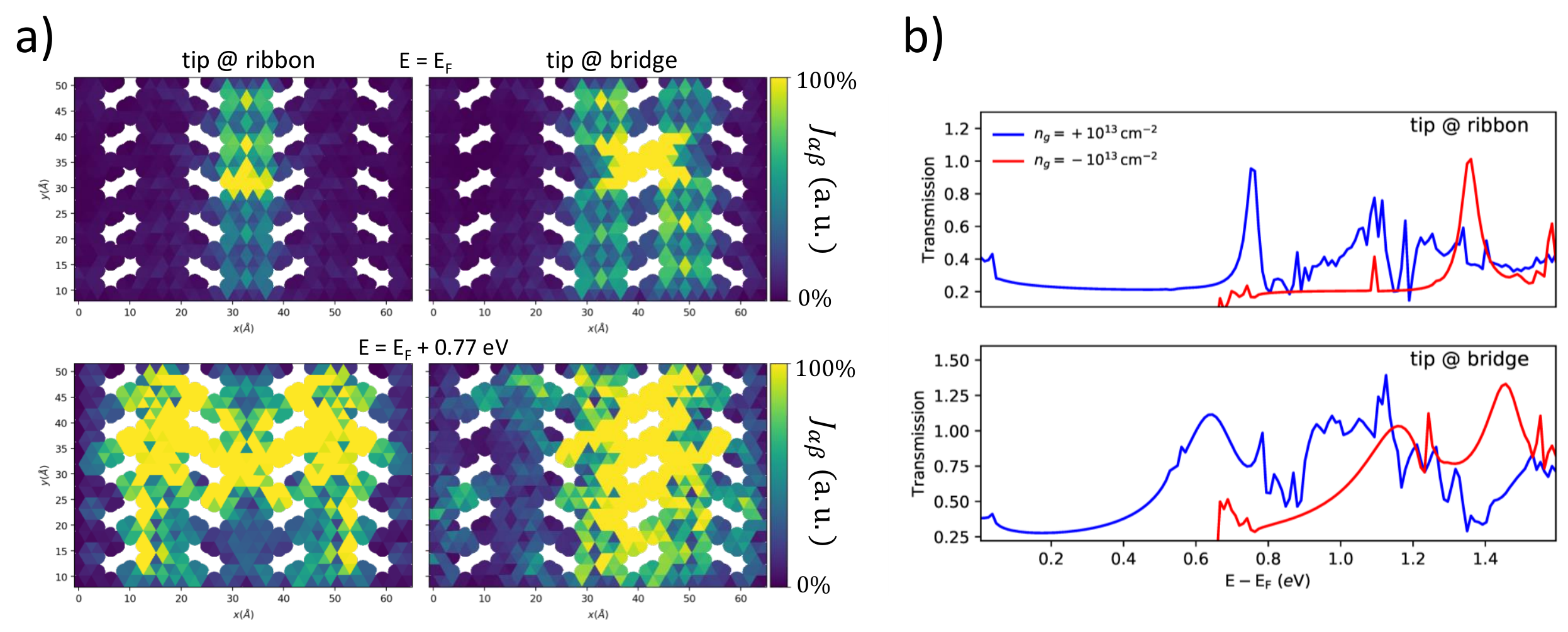}
	\caption{\label{fig:2}
		(a) Bond-currents in the $n$-doped NPG scattering region, injected by the tip while positioned at ribbon and bridge sites, for $n_g=+10^{13}\,\cm^{-2}$. Currents are shown for electrons injected at $E=E_{\rm F}$ or $0.77\,\ev$ above it in the unoccupied bands.
		(b) Transmission spectra for electrons injected by the tip and reaching top and bottom NPG electrodes, for the two tip positions under the two gated conditions. No significant differences between transport towards top and bottom NPG electrodes are observed, so we plot the sum of transmissions towards the two NPG electrodes.
	}
\end{figure*}

In order to visualize the current flow in the 2D mesh we calculate bond-currents among the NPG atoms while injecting electrons with various energies from the tip, as illustrated in \Figref{fig:2}a. We find that, at least up to distances comparable to the cell size, electron injected at the Fermi level or up to $\sim0.7\,\ev$ above it will flow within the same ribbon where they are injected. When the tip is at bridge site the situation is similar but the injected currents propagate in both the two ribbons connected by the probed bridge. At around $0.77\,\ev$ currents start to spread across the bridges and propagate in the neighbor ribbons, as suggested by the marked spike in the transmission of \Figref{fig:2}b. At higher energies crosstalk between neighbor periodic images of the tip makes it hard to visualize the correct current flow in the system.

In \Figref{fig:2}b we plot the transmission spectra for electrons traveling from the gold tip to the two NPG electrodes. For all energies up to $0.7-0.8\,\ev$ above the lowest unoccupied band transmission is quite constant if the tip lies in the middle of a ribbon, whereas it slowly increases when it lies at a bridge site. The onset of transversal bands at higher energies complicates the spectra significantly.

\subsection{Multi-scale method applied to NPG+tip systems}

\begin{figure*}
	\centering
	\includegraphics[width=1.0\textwidth]{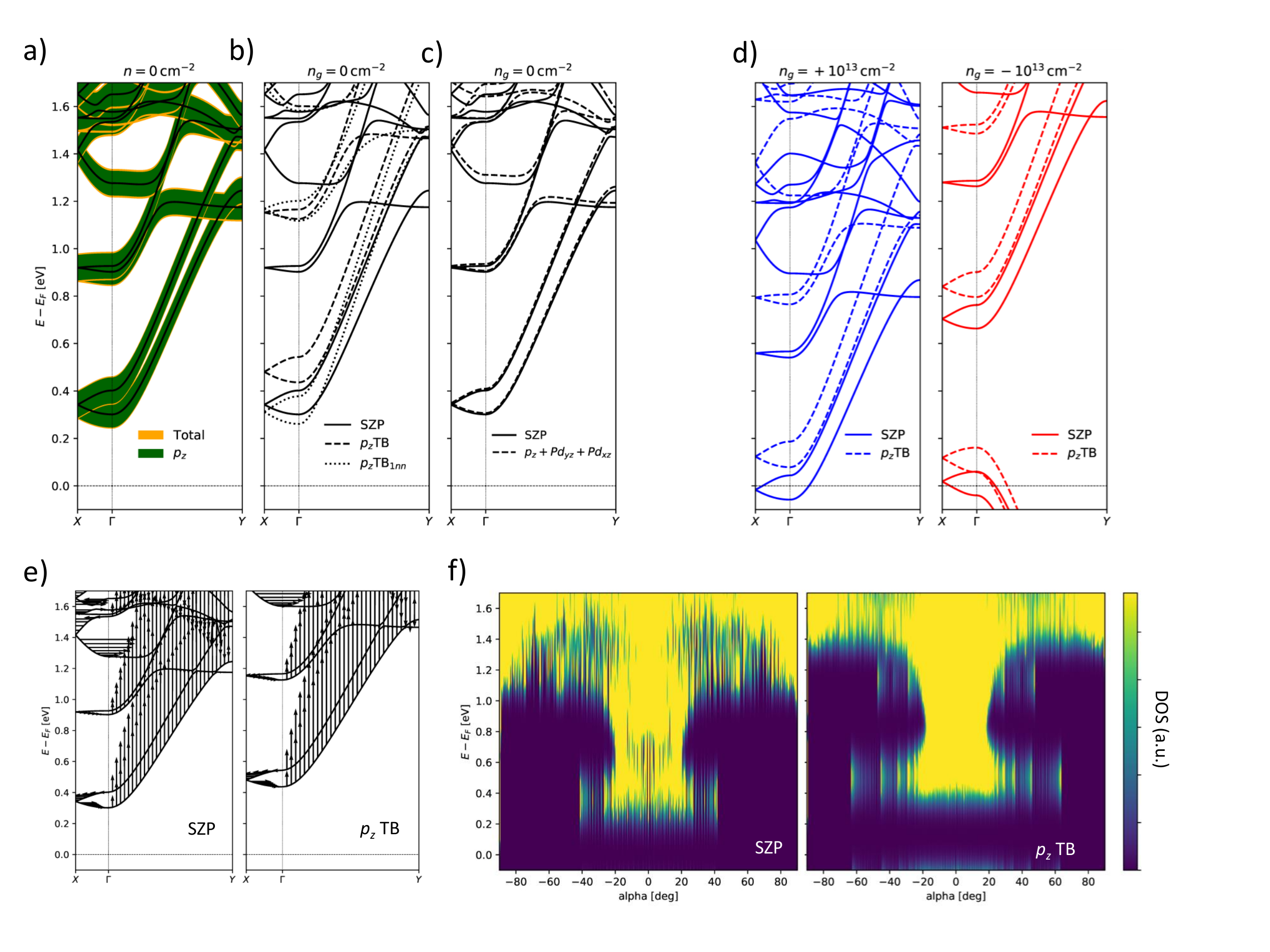}
	\caption{\label{fig:3}
		(a) Fat bands from the DFT model of NPG, showing a predominant contribution of carbons $p_z$ orbitals to the longitudinal bands.
		(b) Comparison between the NPG bandstructure obtained using DFT, a TB model fully parametrized using $p_z$ orbitals from DFT and a nearest neighbor TB model with hopping $t=-2.7\,\ev$. 
		(c) Comparison between the NPG bandstructure obtained using DFT and the TB model fully parametrized using $p_z$, $Pd_{yz}$ and $Pd_{yz}$ orbitals from DFT.
		(d) Comparison between the NPG bandstructure obtained using DFT and the TB model fully parametrized using $p_z$ orbitals from DFT, for different gated conditions.
		(e) Bands velocity vectors at sampled k-points, confirming the predominantly longitudinal or transversal nature of the electronic states within different energy ranges.
		Velocity-angle-resolved density of states for the unit cell of NPG, obtained using the DFT model and the TB model fully parametrized using $p_z$ orbitals from DFT. Maps resolution is converged using a fine grid of k-points.
	}
\end{figure*}

\begin{figure*}
	\centering
	\includegraphics[width=0.7\textwidth]{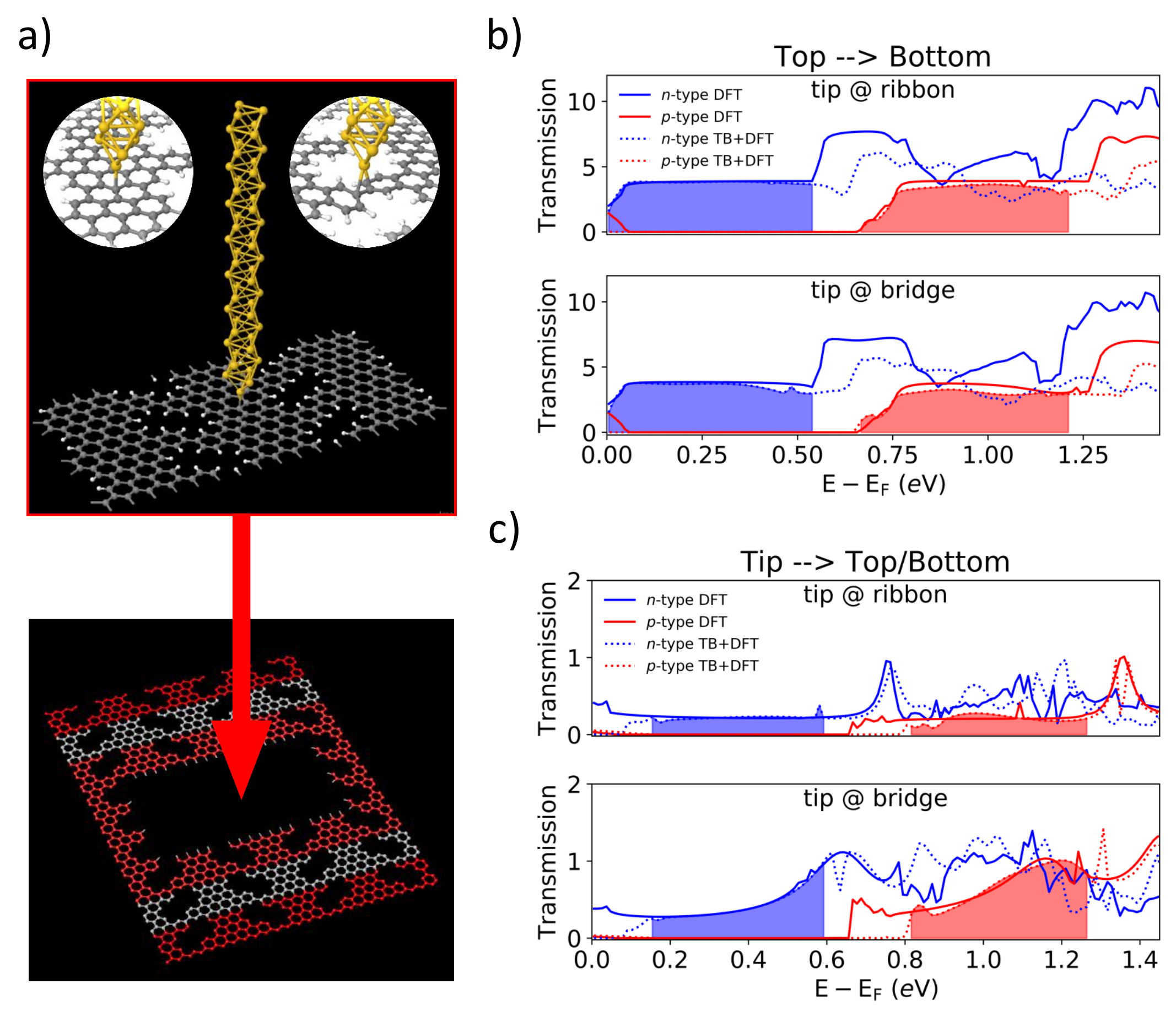}
	\caption{\label{fig:4}
		(a) Schematics of the multi-scale method applied to using the same size and boundary conditions of the DFT model. The perturbed area in the ``guest'' DFT model illustrated on thel left is effectively substituted in the TB model on the right using a self-energy projected on the unperturbed sites around the tip contact region. The red frame-shaped area in the middle contains the self-energy coupling DFT and TB, whereas the two red regions at the top and bottom sides contain the self-energies of two semi-infinite NPG electrodes. The cell has been elongated a little along the semi-infinite direction to ensure no connections among the three electrode regions.
		(b) Transmissions among the two NPG electrodes obtained using the multi-method approach, in comparison to the ones obtained by DFT, for different gates and tip positions. We shift the original curves from DFT+TB by $-0.15\,\ev$ to compensate the orbital-pruning induced offset discussed in \Figref{fig:3}b. The best agreement is found at energies in the shaded range.
		(b) Same as (b) but comparing transmissions from the DFT region to the two NPG electrodes. In this case the original curves from DFT+TB have not been shifted.
	}
\end{figure*}

Below we just highlight a few additional aspects of the multi-scale DFT+TB method applied in the context of this work.

The main idea of the approach is to transfer relevant information about electron transport between two different models by constructing a special self-energy term which connects them. 

The typical protocol applied to study NPG+tip systems, relying on the versatility of \sisl, is: i) setup a DFT model of the contact between NPG and a STM tip; ii) compute the self-energy which connects all C $p_z$ orbitals in the outermost unperturbed area of the DFT cell to all the other orbitals in it; iii) construct a large TB model by projecting a pristine DFT Hamiltonian onto the C $p_z$ orbitals; iv) incorporate the self-energy locally into this TB model as an extra electrode. 

One important requirement is that the DFT cell containing NPG and tip is large enough to have an unperturbed potential in its outermost regions, which is where the self-energy needs to be projected.
Another important requirement is that the TB model is as compatible as possible to the DFT model. This is accomplished by defining the TB parameters directly as the elements in the DFT Hamiltonian and overlap associated to the main orbitals governing transport in the energy range of interest.
For the NPG case, analysis of the various orbital contributions to the DFT bandstructure reveals that longitudinal bands are predominantly dominated by carbons $p_z$ orbitals (\Figref{fig:3}a). The NPG bandstructure obtained with this parametrized TB model is similar to that from DFT, short of some rescaling (\Figref{fig:3}b). For example the TB first unoccupied band is shifted almost rigidly $0.15\,\ev$ above the DFT one.
Compared to a typical nearest-neighbor TB model for graphene with hopping $t=-2.7\,\ev$, the parametrized model used here is non-orthogonal and the range of interaction between TB sites extends as much as in the DFT basis set. Nevertheless such a simplified model also yields bands which compare qualitatively to DFT (\Figref{fig:3}b). Such similarity, although not good enough for embedding a DFT-precision region via a self-energy (which is intrinsically non-orthogonal), suggests that it can be used a simple toy-model for NPG.
The differences between parametrized TB model and DFT can be entirely attributed to pruning of the DFT basis set. In fact, although $p_z$ orbitals almost entirely define DFT bands and govern transport across the system, there are other orbitals in the basis which have a similar symmetry to $p_z$, such as the polarized $d_{xz}$ and $d_{yz}$. These would normally hybridize to some extent and give rise to deviations in the bands. These deviations can be accounted for by incorporating these few extra orbitals into a multi-orbital TB model (\Figref{fig:3}c).
Application of a gate affects the TB bands the same way as the DFT ones (\Figref{fig:3}d).

Analysis of the bands velocity vectors (\Figref{fig:3}e) confirms the predominantly longitudinal or transversal nature of the electronic states as a function of energy, both for the DFT and TB models. Like the current maps reported in \Figref{fig:3}c this seems to corroborate the hypothesis of complete electron confinement within individual ribbons. However a closer look to the density of states in the NPG (see \Figref{fig:3}f) reveals that states in the longitudinal regime \emph{do not} all propagate along the longitudinal direction, but rather their angle of propagation distributed over $\pm 20^\circ$ (regardless the particular DFT or TB nature of the model used to describe them). Such finite angular distribution impacts on the far-field behavior of electrons in NPG.

%%%

We benchmark the applicability of the multi-scale method by inserting the DFT precision injection region inside a geometry which has the same size and boundary conditions of the DFT setup (open along the longitudinal direction and periodic along the transverse) (\Figref{fig:4}a). The results in \Figref{fig:4}b-c show that we can reproduce the DFT transmission among the three electrodes over most of the longitudinal energy range by the combined DFT+TB model within accuracy. In general we find smaller deviations between DFT and DFT+TB when NPG is $n$-doped and when the STM tip is located at a ribbon site. This trend can be attributed to the fact that the bridge site probed by the tip is too close to the outermost region where the connecting self-energy is projected (see \Figref{fig:1}c), thus preventing an optimal match between the outermost DFT potential and the external TB regions. Deviations are expected to be smaller for larger DFT supercell cell sizes.
Conclusively the comparison presented here confirms that the electronic structure of DFT+TB are in very good agreement with the DFT ones, both in terms of bands shape and velocities (see \Figref{fig:1}c), despite the band-misalignment and rescaling caused by orbital-projection of the DFT Hamiltonian (see \Figref{fig:1}c).

\subsection{Transmission and far-field bond-currents from DFT-precision tip to TB-modeled NPG}

\begin{figure*}
	\centering
	\includegraphics[width=1.0\textwidth]{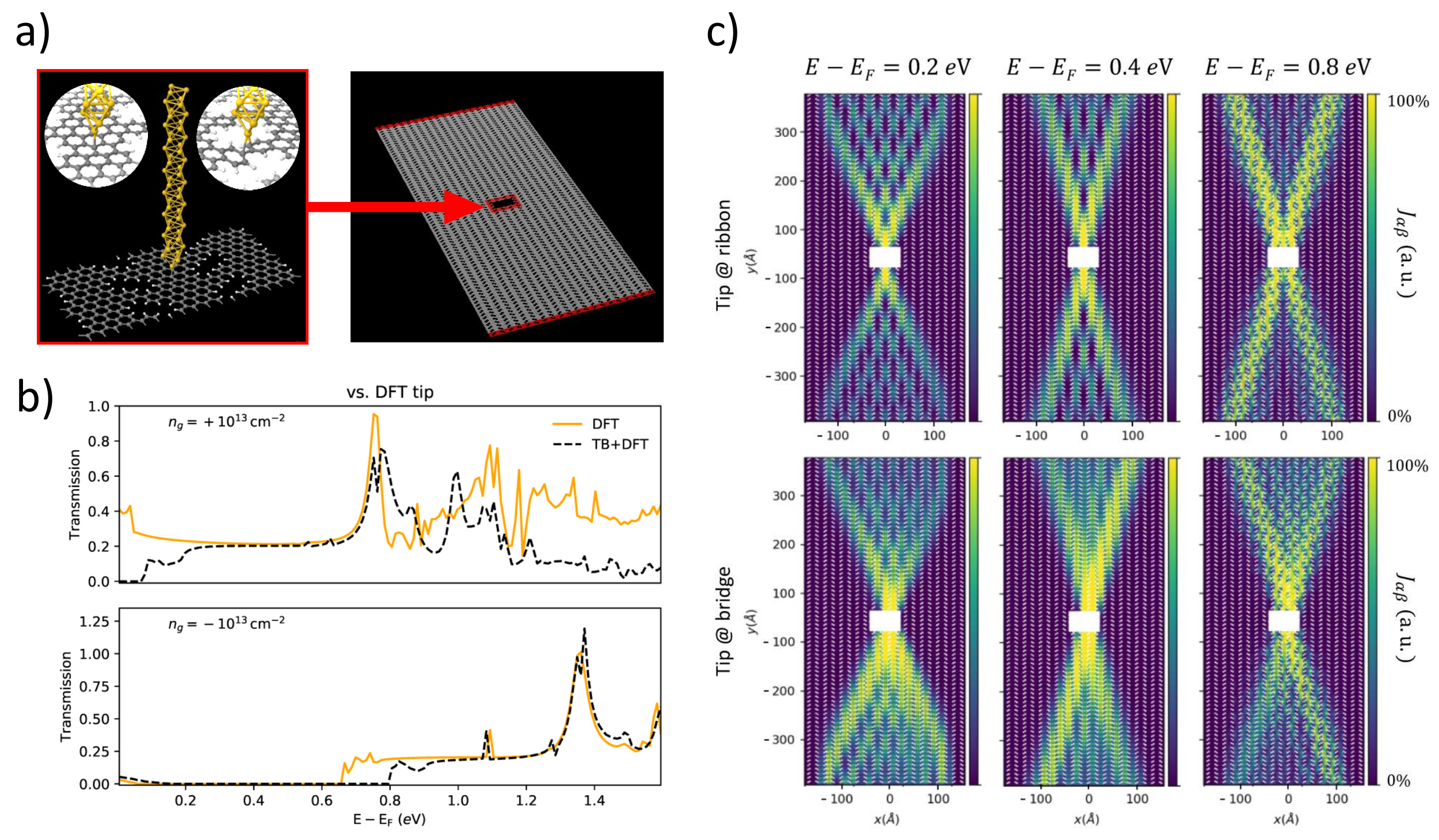}
	\caption{\label{fig:5}
		(a) Schematics of our multi-method multi-scale approach, showing a DFT region with local contact to a gold tip incorporated into a $65\,\nm\times80\,\nm$ TB modeled 2-probe device by means of a coupling self-energy. 
		(b) Total transmission from tip to the two large NPG electrodes obtained using our multi-method multi-scale approach, in comparison to the one obtained by DFT, for the two different gated situations and a tip located at a ribbon site. 
		(c) Dependence of the interference pattern on energy and tip position, for $n_g=+10^{13}\,\cm^{-2}$.
	}
\end{figure*}

We introduce the DFT injection region into a larger $65\,\nm\times80\,\nm$ TB model of a device with 2 bulk NPG electrodes (\Figref{fig:5}a). The transmission spectra from tip to the two large NPG electrodes compare well to DFT (\Figref{fig:5}b), regardless the particular gated conditions or tip position (only ribbon tip position is shown). Deviations can be attributed to the aforementioned pruning-induced bands rescaling.

We find that the interference pattern maximum divergence angle varies slightly with energy (\Figref{fig:5}c). For $n$-doped NPG and tip located on a ribbon, for example, it decreases from $\approx 30\deg$ for $E-E_{\rm F}< 0.3\,\ev$ to $\approx 20\deg$ for $0.4\,\ev < E-E_{\rm F} < 0.8\,\ev$. This is in agreement with the angle-resolved density of states (\Figref{fig:3}f) and the energy dependence of the Talbot coupling strength reported in the main text. At around $0.6-0.7\,\ev$ electrons wavelength becomes comparable to the ribbons width ($w\approx1\,\nm$) hence a finer structure appears. 
Similar results are observed when the injection occurs in proximity of a bridge connecting two neighboring ribbons, although the interference pattern in this case is more blurred. Based on the bond currents of \Figref{fig:2}b, we interpret this as an overlap between currents which are almost equally injected into the two ribbons connected by the two probed bridge atoms.

%\subsection{Bond-current maps from dual-probe calculations}
\begin{figure*}
	\centering
	\includegraphics[width=0.7\textwidth]{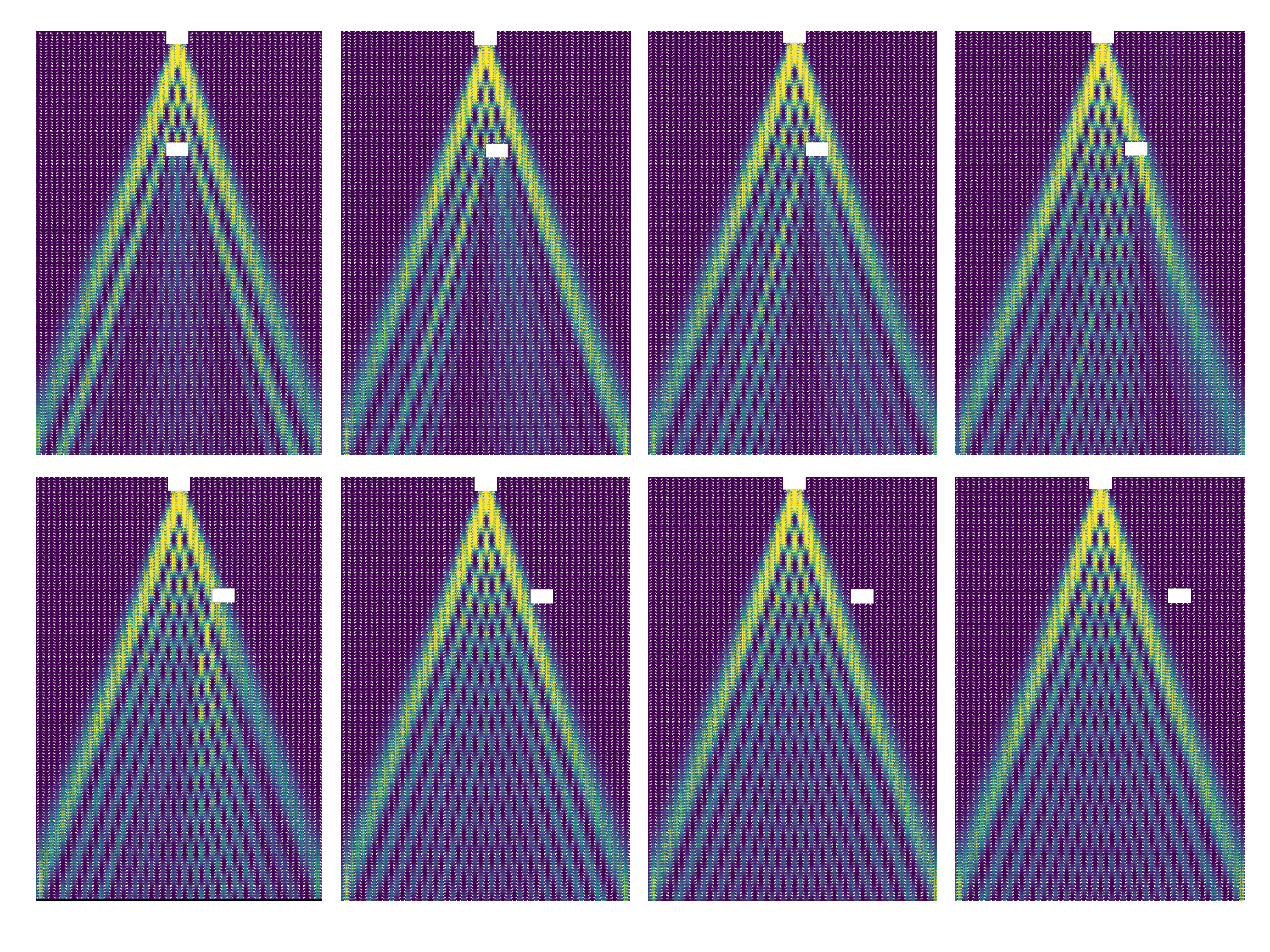}
	\caption{\label{fig:dual2}
		Bond-currents in the NPG with two DFT-precision tips, showing how the second tip, besides detecting, also acts as a significantly strong scattering source. Currents are plotted for $E-E_{\rm F}=0.2\,\ev$ for different positions of the probe tip. Both tips are on top of a ribbon. When the probe tip is on bridges the scattered currents look very similar.
	}
\end{figure*}

%%%%%%%%%%%%%%%%%%%%%%%%%%%%%%%%%%%%%%%%%%%%%%%%%
% Bibliography
%%%%%%%%%%%%%%%%%%%%%%%%%%%%%%%%%%%%%%%%%%%%%%%%%

%\bibliography{tipNPG_bib}


\end{document}